# A Graph-based Method for Session-based Recommendations

MARINA DELIANIDI, MICHAIL SALAMPASIS, KONSTANTINOS DIAMANTARAS, THEODOSIOS SIOMOS, ALKIVIADIS KATSALIS, IPHIGENIA KARAVELI

All authors are affiliated with the International Hellenic University, Thessaloniki, Greece

We present a graph-based approach for the data management tasks and the efficient operation of a system for session-based next-item recommendations. The proposed method can collect data continuously and incrementally from an ecommerce web site, thus seemingly prepare the necessary data infrastructure for the recommendation algorithm to operate without any excessive training phase. Our work aims at developing a recommender method that represents a balance between data processing and management efficiency requirements and the effectiveness of the recommendations produced. We use the Neo4j graph database to implement a prototype of such a system. Furthermore, we use an industry dataset corresponding to a typical e-commerce session-based scenario, and we report on experiments using our graph-based approach and other state-of-the-art machine learning and deep learning methods.

**CCS CONCEPTS • Computing methodologies**

**Additional Keywords and Phrases:** Neo4j graph database, eCommerce, Session-based recommendations

## 1 INTRODUCTION

Recommendation Systems (RS) make personalized recommendations to ecommerce users for additional products that might be of interest to them. These systems find similar items, by collecting and modeling previous user transactions, as well as other features such as location, user demographic profile and other people's preferences. They finally provide, for each targeted user action, a list of top N recommended items, occasionally after applying desired business rules. In RS research approaches that utilize long-term user profiles are predominant [7]. However, in many applications (e.g. ecommerce), such long-term user models are often not available for privacy reasons. Recommendation techniques which rely only on the user's recent behavior and other session-specific data, and which adapt their recommendations to the recent user's actions, are called session-based recommendation approaches (SBRS). Their main objective is, for each viewed item, to recommend a set of the next-items in every ongoing user session.

Early recommendation methods to predict the next-item were based on sequential pattern mining techniques using relatively simple statistical co-occurrence analysis. More sophisticated methods based on Context Trees [8], Markov models [9], Reinforcement Learning [10] and more recently Recurrent Neural Networks [11] methods (RNNs) were tested with very good results. The main advantage of the simpler techniques based on frequent patterns analysis was that they are easy to implement and lead to straightforward interpretable models. They perform good and robust in many applications without training and maintenance costs, but generally RNNs perform better. Regardless of the method applied, the mining process in RS is computationally demanding.

Running a RS is a data intensive operation and an abundance of information must be collected and stored something which is usually done in relational databases. Then, the data are processed and analyzed to create prediction models and also to facilitate data provision to the recommender component. Furthermore, the exponential growth of the web led to the explosion of user data that can be collected from multiple channels, resulting in data management challenges for RSs. Similarly, the processing requirements have been also increased in order to produce effective recommendations in a more complex environment.

This is precisely the need that we address using a graph-based approach for the data management tasks and the development of an RS. Specifically, we aim at developing a method that will represent a balance between the data processing and management requirements and on the other side the effectiveness of the recommendations produced. Our method is to develop a simple and fast method of making recommendations using a graph database. The main idea is to trace item pair co-occurrences during each session using a graph. By counting the number of co-occurrences of item pairs, we calculate the degree of similarity of each pair. This operation creates a relevance table which is subsequently used to produce the list of recommended items at every user action as the session continues.

The tool that we use to implement our method is the Neo4j Graph Database (GDB). Neo4j is an open-source NoSQL graph database implemented in Java that stores data as graphs. In our specific application it stores all session data (items and connections between them) as a connected graph. Neo4j provides full database capabilities such as ACID (atomicity, consistency, isolation, durability) transaction support, cluster support, backup and redirection, and so on. The basic modeling in Neo4j consists of nodes, properties, and relationships. Nodes represent objects and connect to each other by linked edges. An edge represents the relationship between objects. Node and edge can have properties that indicate the attributes of the object/relationship. It only needs to know Cypher, a query language similar to SQL, to use Neo4j and do not need to delve deep into graph theory. Complex data management is implemented using queries that are intuitive and simple for the developer. In addition, Neo4j can be used in many programming languages that have interface class libraries [1]. According to [2], Neo4j has a powerful traversal framework and query languages for traversing the graph and can be deployed as a standalone server or an embedded database with a very small distribution footprint.

Our graph-based method is, to some extent, similar to other recommendation systems that are implemented using graph models. Content-based, collaborative filtering or rule-based recommendation systems were also developed using graphs. User-based collaborative filtering for movie recommendations employing graphs has been presented in [1]. It improves the user experience despite the sparsity of the data and speeds up the extraction of the user recommendations list. An example of using graphs for social recommendations is presented in [5]. The authors apply a nodes' transitivity algorithm for social recommendations and improve the trust between users by using social tagging. In [6] the authors use a graph database for the ontology transformation and then, using rule-based recommendations and RFM (recency, frequency, monetary) analysis for customer behavioral knowledge, make personal goods recommendation lists. A content-based filtering recommendation approach using Neo4j [4] proposes a search of article abstracts based on document-keyword relation. The graph was used to filter keyword-document co-occurrence as a similarity weight between documents in order to reduce searching space and recommend relative documents our graph-based method differs from all these previous methods in the fact that we use session data to recommend items for the next session step based on simple co-occurrence statistics.

The remainder of the paper is structured as follows. Section 2 presents the implementation details of our graph-based recommendation method using the Neo4j graph database. The experiment is described in section 3. Conclusions and feature work are discussed in Section 4.

## 2 A GRAPH-BASED RECCOMENDER SYSTEM

### 2.1 Data modeling

Data modeling using graphs defines nodes and relations between them. The existence of relations in almost any information problem suggests that almost anything can be modelled as a graph.

In our case, we consider a collection of session data from an e-commerce web site. Every step in a session contains information about the session-id, the product-id, the category of the product and the user action taken. No information is available about the anonymous user. In order to describe the data in a graph structure we use the Neo4j graph database. In our design, the graph consists of three types of nodes and four types of relations between the nodes (see Figure 1). The three types of nodes represent *Products, Sessions* and product *Categories*. There are 4 relations that connect the nodes with each other. *ProductInSession* relation shows all the products that a user interacts in a session. The *IsParentOf* relation connects the *Category* nodes between them making a tree structure with five levels depth while the *BelongsTo* relation connects the products with the categories their belong to. The nodes are described with attributes such as session id (*sid*), product id (*pid*), category id (*catid*). The relation *ProductInSession* is uniquely identified by the attributes *step* that shows the ordered time sequence appearance of a viewed product during the session, and *action_id*, *action* for the kind of action at the step in the session. The *InSameSession* relation connects pairs of products that appear together in some session. The *InSameSession* attribute *popularity* counts how many times a pair of products appears in all sessions and represents the co-occurrence of a product pair. A product in a session can be paired with another product or with itself. The *InSameSession* relation and the *popularity* attribute are pivotal for the implementation of our recommendation algorithm.

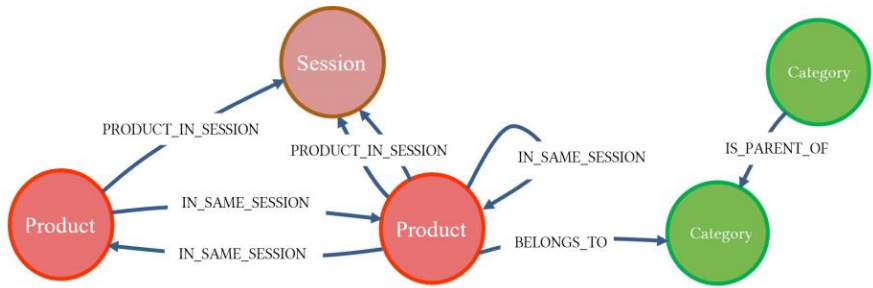

Figure 1. The Neo4j Graph Database model of the e-commerce company

### 2.2 Recommendation algorithm

We work on the assumption that if product A has high co-occurrence with product B, then B is a potential recommendation if the user observes A. Traversing the graph through the *InSameSession* relation we calculate the frequency of product pairs co-occurrence for all sessions and find the degree of similarity between the products. The product pairs are then sorted by the similarity degree in descending order and stored in a table.

Assuming that, during a session, the user observes product A in step i, we recommend the top-k products most similar to A according to the sorted list in the table. We measure the effectiveness of our algorithm using the Mean Reciprocal Rank (MRR) metric over all recommendations of all steps in all sessions, except for the last step in every session because it is not possible to evaluate our recommendation on the last item of a session.

## 3 EXPERIMENT AND RESULTS

### 3.1 Dataset

The raw data for the experiments related to the next-item task are taken from the web server logs of an e-commerce application (leather apparels) for a relatively long period of time (eight months). The log data were analyzed to identify sessions, session length, user actions in each session, actions' related items, item categories, and time spent in each action. The user-agent cookie of each log line is used to identify unique sessions. Each sequence identified by a unique cookie is a user's session and consists of the actions that the user has made during a session. The above dataset was processed to obtain only the sessions that contain at least 2 behavior sequences. As a result of this preprocess the dataset consists of 24111 sessions that altogether count 312912 user actions. The complete set of action types were retrieved for website user behavior analysis (click-stream analysis), but this work will be reported on another paper. The 728 sessions that ended in purchases means a conversion rate of 3%. In 91.2% of sessions (i.e., 22.008 sessions), users did not have any items in their shopping cart when they exited, which implies sessions that were pure browsing only. The rest of the sessions had items in their shopping cart when they finished, but never turned into purchases. Twelve different action types were identified out of the log files (e.g. view product, add-item-to-cart, checkout etc.). For the next-item experiment which is presented in this paper, only the view item user actions were considered (i.e. View Product) and only the sessions that include at least two different item views, finally pertaining to a number of 12128 applicable sessions consisting of 67101 user View Product actions.

### 3.2 Metric

When running our experiments, for each test session we iterate the items sequence and for every viewed item in the sequence we produce a set of recommended items using all methods. After obtaining a ranked list of n recommended item(s) we calculate the Reciprocal Rank for each visited item from the session sequence. When all items are considered (except the first one of each session), we calculate the Mean Reciprocal Rank (MRR) of the entire session and then again, we average the MRR of all tested sessions to calculate the MRR of the method and these are the figures that are reported in this Section. MRR is a statistic measure for evaluating any process that produces a list of possible responses to a sample of recommendations. The reason we choose MRR as the evaluation measure is because it expresses the effectiveness of a method to recommend the next-item as higher as possible in the recommendation list. The general underlying assumption is that if a method attains an MRR between 0.20 to 0.25, then a system would require a list of 4-5 recommended items to effectively predict the next viewed item.

### 3.3 Experiment

In the experiment that we conducted, 90% of the dataset was used for the train set (60815 samples) and the remaining 10% for the test set (6286 samples).

To compare our graph-based recommender method we have used the same dataset and tested other more complex recommender methods, namely: ItemToVec, Doc2Vec and LSTM. These implementations and experiments are reported in detail in another paper [12] however in this paper we include their results in order to present a complete comparison of our graph-base method to other well know methods for session-based recommendations. As we earlier discussed, the advantage of our graph-based recommendation method is it's easier data management, no training is required from time to time, and the efficiency of the recommendations relates to the efficiency of the graph database tool which is quite stable. However, our work on comparing our method to RNN and embeddings methods for next-item session-based recommendations was mainly driven by the need of a more complete and systematic comparison of that type. We generally know that simple frequency pattern analysis methods like the one we have implemented were quite effective in the past. However, to what degree that observation would be repeated with our graph-based method was unknown.

### 3.4 Results and discussion

The products pair popularity or co-occurrence can be used as a similarity score for product recommendation in the session-based dataset we worked.

We calculate the MRR in 2 ways. As we mentioned, a product can appear in the same session more than 1 times. We called that pairs as self-pair popularity. So, the MRR measure has been calculated twice, with and without product's self-pair popularity. As can we see, our proposed method performs much better than the relatively static methods based on embeddings attaining and MRR of 0.170 which is significantly larger than the Item2vec and the Doc2Vec methods. However, it is signciantly outperformed by the LSTM method which is of a more dynamic nature but requires a lot of training.

Table 1: MRR results of all recommender methods

| Method | **MRR** |
| --- | --- |
| Graph-based (self-pair excluded) | 0.170 |
| Graph-based (self-pair included) | 0.165 |
| Doc2Vec | 0.101 |
| Item2Vec | 0.087 |
| LSTM | 0.265 |

### 4 CONCLUSION AND FUTURE WORK

Recommender systems help customers to discover products they need growing customer satisfaction and creating a better shopping experience. It is therefore not a surprise that RS play an important role supporting e-commerce applications to increase customers' basket size and sales figures. Also, another important functional requirement for RS is to be able to accommodate business rules that are desired for the management and the everyday operation of an e-commerce application. An example of such business rule is the promotion of items from specific categories for a period of time, or to satisfy item coverage constraints, i.e. every item is recommended at least to a certain number of users. All these examples make clear that designing a RS is a multi-faceted and multi-criteria problem. In other words, recommendations must be done effectively to suggest relevant next-item as high as possible in the recommendation list, but at the same time the complete operation cycle of the RS (data gathering, modelling, processing, analysis, filtering) should be executed efficiently and flexibly.

Our work on developing a graph-based system for session-based next-item recommendations was mainly driven by the ideas discussed above. The proposed method has some important properties that address several of the abovementioned needs. It can collect data continuously from an ecommerce web site from all ongoing user sessions and immediately make them available to the recommendation algorithm without any training phase. Also, because it is developed on a flexible and powerful graph database platform, it is flexible to integrate new business rules and constraints on demand, something that will be very difficult to inherently implement using ML methods.

Generally, scalability is a critical issue that should be very carefully considered, especially in large scale applications [13]. For example, a method achieving very high effectiveness, but it requires a lot of training time exceeding the time that the recommender prediction model should be periodically updated (e.g. every day or every week), it is not practical and not applicable.

In conclusion we feel that our work in developing our graph-based recommender method has produced some useful insights into the development of session-based recommenders which will help engineers to produce well -balanced (i.e. efficiency vs effectiveness) recommender components for ecommerce and other session-based applications.

## ACKNOWLEDGMENTS

This research has been co-financed by the European Regional Development Fund of the European Union and Greek national funds through the Operational Program Competitiveness, Entrepreneurship and Innovation, under the call RESEARCH–CREATE–INNOVATE (project code:T1EDK-01776)